\documentclass{aa}  
\usepackage{graphicx}
\usepackage{txfonts}
\usepackage{fixltx2e}

\usepackage{textcomp,gensymb}
\usepackage{hyperref}
\hypersetup{
  colorlinks=true,
  citecolor=blue,
}
\usepackage{multirow}

\newcommand{\ugamma}{J\,K\textsuperscript{-1}\,m\textsuperscript{-2}\,s\textsuperscript{-0.5}\,}
\newcommand{\urho}{g\,cm\textsuperscript{-3}\,}
\newcommand{\uvel}{km\,s\textsuperscript{-1}\,}

\begin{document} 

\title{Influence of the Yarkovsky force on Jupiter Trojan asteroids}


\author{S. Hellmich
  \inst{1}
  \and
  S. Mottola\inst{1}
  \and
  G. Hahn\inst{1}
  \and
  E. K\"uhrt\inst{1}
  \and
  D. de Niem\inst{1}
}

\institute{Institute of Planetary Research, German Aerospace Center (DLR),
  Rutherfordstr 2, 12489 Berlin\\
  \email{stephan.hellmich@dlr.de}
             }


 
  \abstract
  {}
   {
     We investigate the influence of the Yarkovsky force on the long-term orbital evolution of Jupiter Trojan asteroids.
   }
   {
     Clones of the observed population with different sizes and different thermal properties were numerically integrated for 1 Gyr with and without the Yarkovsky effect. The escape rate of these objects from the Trojan region as well as changes in the libration amplitude, eccentricity, and inclination were used as a metric of the strength of the Yarkovsky effect on the Trojan orbits.
   }
   {
    Objects with radii $R\leq$1 km are significantly influenced by the Yarkovsky force. The effect causes a depletion of these objects over timescales of a few hundred million years. As a consequence, we expect the size-frequency distribution of small Trojans to show a shallower slope than that of the currently observable  population ($R$ $\gtrsim$ 1 km), with a turning point between $R$ = 100 m and {$R$ = 1 km}. The effect of the Yarkovsky acceleration on the orbits of Trojans depends on the sense of rotation in a complex way. The libration amplitude of prograde rotators decreases with time while the eccentricity increases. Retrograde rotators experience the opposite effect, which results in retrograde rotators being ejected faster from the 1:1 resonance region. Furthermore, for objects affected by the Yarkovsky force, we find indications that the effect tends to smooth out the differences in the orbital distribution between the two clouds.
   }
   {}


\maketitle
%

\section{Introduction}

Jupiter shares its orbit with a large number of asteroids, called
Jupiter Trojan asteroids, that populate the regions around the
L\textsubscript{4} and L\textsubscript{5} Lagrangian points in the
Sun-Jupiter system. It is still debated how these objects were
emplaced into their current orbits. However, it was demonstrated that
a large part of the known population is indeed stable over the age of
the solar system \citep{erdi_long_1988} and detailed maps describing
the stable regions around the Lagrangian points have been created
\citep{levison_dynamical_1997, tsiganis_chaotic_2005}.  An interesting
peculiarity of the Jupiter Trojans is that the two swarms
significantly differ in the number of objects. As of August 2019, the
Minor Planet Center lists 4603 numbered, multi-, and single-opposition
Trojans in the leading (around L\textsubscript{4}) but only 2476 in
the trailing cloud (around L\textsubscript{5}). It can be excluded
that this asymmetry originates from an observational bias
\citep{grav_wise/neowise_2011}. Although the gravitational stability for
orbits around either of the Lagrangian points is similar
\citep{marzari_instability_2002}, it was found in long-term
integrations of the known Trojan population that the fraction of
escaping particles from the L\textsubscript{5} cloud is larger than
that from L\textsubscript{4} \citep{di_sisto_dynamical_2019}. However,
\cite{di_sisto_dynamical_2019} also concluded that the difference in
escaping objects alone cannot explain the asymmetry in the Trojan
clouds that is observed today. A theory capable of describing both
orbital distribution and asymmetry is the \textit{\textup{jump capture}}
scenario proposed by \cite{nesvorny_capture_2013}. According to this
model, the Trojans originate from more distant objects, initially
located between 5 and 30 au, which were scattered inward during a
phase of orbital chaos in the early solar system and eventually became
trapped in resonance when Jupiter and Saturn reached their current
orbits. In this model, the asymmetry is explained by the influence of
Uranus, Neptune, or a hypothetical ice giant (now ejected from the
solar system) that may have moved through one of the Trojan clouds
and wiped out a large number of objects. By simulating the early phase of
the solar system, \cite{pirani_consequences_2019} recently showed that
particles in the gaseous protoplanetary disk can be captured around
the Lagrange points of Jupiter while the growing planets migrate
through the disk. Because the drift is directed inward, the process of
particle capture naturally favors tadpole orbits around
L\textsubscript{4}, and the resulting asymmetry after migration and
planet formation is compatible with the observed asymmetry. However, their model fails to explain the inclination
distribution of the known Trojans.

Only very little is known about the physical properties of the Jupiter
Trojans.  Using infrared observations carried out with the Wide-field
Infrared Survey Explorer (WISE) space telescope, a low mean geometric
albedo of 0.07$\pm$0.03 was determined for a sample of about 3000
objects \citep{grav_wise/neowise_2011}. Density estimates exist for
only two Jupiter Trojans: for (624) Hektor and (617)
Patroclus. For (617) Patroclus, several studies report low bulk
densities ranging from 0.8 \urho \citep{marchis_low_2006} to 1.3 \urho
\citep{merline_asteroids_2002}, with 0.88 \urho \citep{buie_size_2015}
being the most recent result. Densities found for (624) Hektor vary
from 1.0 \urho \citep{marchis_puzzling_2014} to 2.48 \urho
\citep{lacerda_densities_2007}. Thermal properties are also measured
for only a few Trojans. \cite{ferrari_thermal_2018} suggested that
thermal inertia decreases with increasing heliocentric distance
and reported a thermal inertia of 10 \ugamma and lower for objects
beyond Saturn, to which the Trojans might be linked due to their dynamical history.  In general, it is expected that Jupiter Trojans
have very low thermal inertia of about 5 \ugamma
\citep{dotto_troianis:_2008}. Using thermo-physical modeling from
infrared observation data, a thermal inertia of 6$^{+4}_{-6}$ \ugamma
has been determined for (624) Hektor \citep{hanus_thermophysical_2015},
while for (1173) Anchises, a higher value in the range of 25 to 100
\ugamma was reported \citep{horner_1173_2012}. By examining
temperature changes on the surface of (617) Patroclus during mutual
events caused by its moon Menoetius, a thermal inertia of 20$\pm$15
\ugamma was determined
\citep{mueller_eclipsing_2010}. \cite{fernandez_albedo_2003} placed
upper limits on the thermal inertia for (2363) Cebriones and (3063)
Makhaon of 14 and 30 \ugamma, respectively.

The stability regions around the L\textsubscript{4} and
L\textsubscript{5} Lagrangian points of the Sun-Jupiter system are
well known. The main parameter controlling the orbital stability of
Jupiter Trojans is their libration amplitude $D$ , which defines the
maximum excursion in mean longitude $\lambda$ with respect to Jupiter
\citep{tsiganis_chaotic_2005},
\begin{equation}
\lambda - \lambda_J = \pm \pi / 3+ D \cos\theta + \mathcal{O}({D}^2),
\label{equ:libration_lon}
\end{equation}
where $\theta$ is the phase of libration, as well as their
eccentricity and inclination. For L\textsubscript{4} Trojans, the
difference in mean longitude is positive, while it is negative for
L\textsubscript{5} Trojans. \cite{tsiganis_chaotic_2005} characterized
the residence time of objects around the Lagrangian points of Jupiter
as a function of libration amplitude, eccentricity, and
inclination. They provided detailed stability maps and found that in
general, the residence time increases with decreasing libration
amplitude and eccentricity. Inclination affects the long-term behavior
to a lesser degree (see Figures 2-5 in
\citealt{tsiganis_chaotic_2005}).

The Yarkovsky effect describes a nongravitational force that is
caused by diurnal heating and asymmetric reradiation of energy
received from the Sun \citep{hartmann_reviewing_1999}. Depending on
the asteroid's rotational parameters and the physical properties of
its surface material, the effect can lead to a secular change in the
orbital semimajor axis. In the main belt, the Yarkovsky effect is the
dominant mechanism that drives asteroids toward unstable resonance
regions where they are eventually ejected from the belt
\citep{farinella_meteorite_1998}. It also causes dynamical spreading
of collisional asteroid families \citep{bottke_dynamical_2001}.

Many studies have been carried out to answer the questions about the
origin and dynamical evolution of Jupiter Trojans. Some authors argued
that due to the large heliocentric distance of Jupiter Trojans, the
Yarkovsky force is too weak to remove them from the stable region
inside the resonance \citep{yoshida_size_2005,
  di_sisto_giga-year_2014}. Others concluded that at least for the
smaller bodies, the Yarkovsky force may become important for their
long-term orbital evolution
\citep{tsiganis_chaotic_2005,karlsson_creation_2011}. So far, the only
orbital evolution studies of Trojans that incorporated the Yarkovsky
effect were performed by \cite{wang_dynamical_2017} and
\cite{hou_dynamics_2016}. They used a simplified model for the
Yarkovsky force and demonstrated in numerical experiments considering
the Sun, Jupiter, Saturn, and a single meter-sized Trojan asteroid that
the Yarkovsky force causes the oscillation in semimajor axis to
increase, with the object escaping the Trojan region after a few tens
of million years.

This study is intended to address the question of the size at which the
Yarkovsky force starts to affect the population of Jupiter Trojans. We also study
whether it is able to significantly influence their orbits on timescales of hundreds of million years.

\section{Methods}
  
\subsection{Creating a synthetic population of Trojans}
\label{sec:a_synth}

For the purpose of explaining certain dynamical properties of the
Trojans such as the difference in escaping particles from each cloud,
the population of currently known Trojans is too small.  Moreover, some
objects have been observed during a single opposition only, which
means that their orbits may not be known with sufficient
accuracy. Therefore, when the long-term evolution is investigated, only the
orbital elements of the numbered and unnumbered multi-opposition
Trojans should be used, which shrinks the set of available orbits even
more. Considering only those particles for our numerical study
would not allow us to draw statistically meaningful conclusions.
Therefore, a large synthetic population of Trojans was created by
cloning the orbits of the numbered and unnumbered multi-opposition
Trojans. Because the long-term stability of the Trojans is very
sensitive to their orbital distribution, care should be taken that
cloning does not change the long-term dynamical behavior of the orbits.  On
that account, the variations in orbital elements for each clone were
computed based on the accuracy of the astrometric positions from which
the orbit of the real object is calculated. The uncertainties for each
orbit are published in form of a covariance matrix on the AstDyS
website\footnote{\href{https://newton.spacedys.com/astdys2/index.php?pc=4}{https://newton.spacedys.com/astdys2/index.php?pc=4}}. Using
that covariance matrix, we can express the uncertainty $\Delta \mathbf{q}$ in the
orbital elements vector $\mathbf{q}$ as
\begin{equation}
  \Delta \mathbf{q} = \underset{i=1}{\overset{6}{\sum}}
  \xi_i\sqrt{\lambda_i}\mathbf{X}_i,
  \label{equ:cov_clones}
\end{equation}
where $\xi_i$ are Gaussian-distributed random variables with a
standard deviation of unity and zero mean, $\lambda_i$ are the
eigenvalues of the covariance matrix, and $\mathbf{X}_i$ are the
normalized eigenvectors. This cloning technique was used to generate a
total number of 98304 Trojan particles. Because only precisely
determined orbits were chosen for cloning and each clone is a
realization of its parent orbit within the 1$\sigma$ range of the
observational data, the long-term orbital evolution of the cloned
population is expected to be representative of the current population.

The clones were generated from numbered and multi-opposition Trojan
orbits, consisting of 3642 L\textsubscript{4} and 1925
L\textsubscript{5} objects. In order to create a population that
contains an equal amount of L\textsubscript{4} and L\textsubscript{5}
Trojans, the cloned population incorporates the real orbits plus 13 -
14 clones for L\textsubscript{4} Trojans and 25 - 26 clones for
L\textsubscript{5} Trojans. For efficiency reasons, the
population of clones was split into groups of 32768 particles, and we started an
individual simulation for each group. In order to determine
whether the Yarkovsky force influences the long-term orbital behavior
of this population, we conducted simulations with and without the Yarkovsky effect. To provide a good baseline against
which the results from the Yarkovsky run could be compared with, the
full sample of 98304 particles was used for the non-Yarkovsky run,
while for the Yarkovsky run a smaller number of 32768 particles was
simulated. For the Yarkovsky run, the subpopulations of the two
clouds were further subdivided into 16 groups of particles with
different physical properties and sizes each. We defined four size
classes with radii of 10 m, 100 m, 1 km, and 10 km. The object density and thermal inertia were split into two
categories each, set to the upper and lower bounds of the values
reported in the literature (see Table \ref{tab:props_trojans}). This
subdivision resulted in 1024 particles for each category, size, and
cloud. The spin-axis orientation was uniformly distributed over the
celestial sphere and the spin period (in seconds) was set to five times
the body's radius (in meters)
\citep{farinella_semimajor_1999}. Geometric albedo ($p_v$) and
infrared emissivity ($\epsilon$) for all objects were fixed to 0.07
and 0.9, respectively \citep{grav_wise/neowise_2011}. To compute the
Bond albedo ($A$), which is relevant for the Yarkovsky force, the
relation $A \approx q p_v$ with the phase integral $q=0.290+0.68G$
\citep{bowell_Application_1989} and a slope parameter $G$ = 0.15 was
used.

We caution that the orbital distribution of the cloned
population is based on the currently known population. This consists
of large objects and might be different for the actual population of
Trojans that are smaller than a few kilometer in radius.

\begin{table}
\centering
\caption{Combinations of thermo-physical properties used for the Yarkovsky effect.}
\begin{tabular}{ccccc}
\hline
Category & $\rho$ [\urho] & $\Gamma$ [\ugamma]& $A$ & $\epsilon$ \\
\hline
1 & 0.8 & 10 & & \\
2 & 0.8 & 100 & &  \\
3 & 2.5 & 10 & & \\
4 & 2.5 & 100 & \multirow{-4}{*}{0.027} & \multirow{-4}{*}{0.9} \\
\hline
\end{tabular}
\label{tab:props_trojans}
\end{table}

\subsection{Collisional lifetime of the Jupiter Trojans}

On the one hand, under the gravitational influence alone, the orbits
of a large number of Jupiter Trojan asteroids can survive the age of
the solar system (about 4.5 Gyr). On the other hand, the lifetime of
asteroids is also limited by mutual collisions that are not resolved
by the integration method used in this work. Therefore, the
integration interval should not exceed the expected collisional
lifetime of the involved bodies. To choose a
realistic integration interval, we therefore estimated the collisional lifetime
of Jupiter Trojans.

Here, the collisional lifetime of a minor body is considered as the
mean time until the object is destroyed by a catastrophic disruption. Such an event is defined as a collision in which the size of the largest remnant is smaller than or equal to half the size of the original body, and the fragments
are dispersed \citep{benz_catastrophic_1999}. The relevant
parameters for the collisional lifetime are size and density of the
colliding bodies as well as the intrinsic collision probability
$P_{i}$, average impact velocity $v_{i}$, and size distribution of the
collisional interacting population. The physical parameters
and impact velocity of the bodies are relevant for determining the energy per unit mass
that causes a catastrophic disruption of the target body ($Q^*_D$). The
intrinsic collision probability, which defines the flux of impactors
per area and time, is a geometric property, and can be combined with
the object size distribution to estimate how frequently catastrophic
disruption events occur.

\begin{figure*}
\centering
\includegraphics[width=1.0\textwidth]{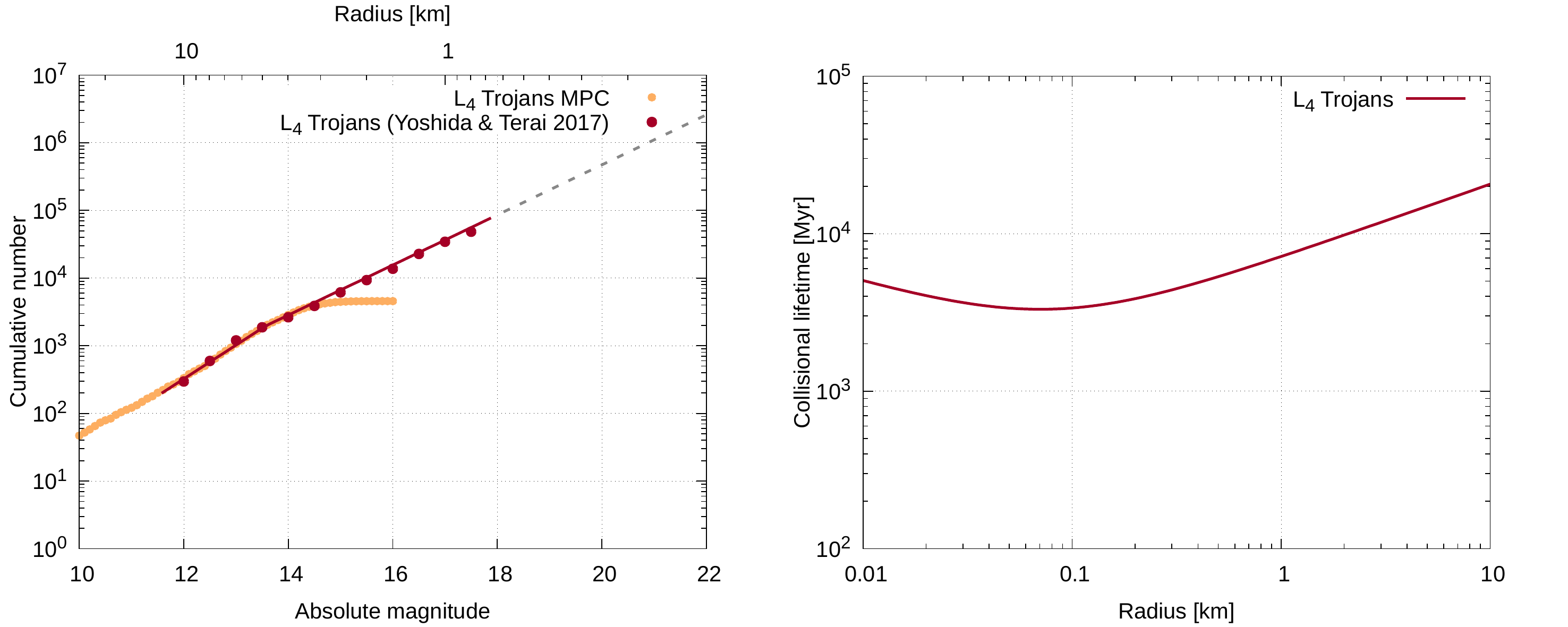}
\caption{Left panel: Cumulative size distribution of L\textsubscript{4} Trojans. Orange: known population; red dots: Trojan population after \cite{yoshida_small_2017}; red line: Broken power law \cite{yoshida_small_2017} fit to their observations; dashed gray line: Extrapolation of the power law down to $R=100$ m objects. Right panel: Collisional lifetime of L\textsubscript{4} Trojans according to this work.}
\label{fig:trojan_size_tau}
\end{figure*}

The average impact velocities and the intrinsic collision probability
for Jupiter Trojans have been determined in several studies
\citep{marzari_collision_1996,delloro_trojan_1998}. However, providing
a good estimate for $Q^*_D$ is not easy. By simulating collisions
using hydrocode and performing high-velocity impact experiments,
several methods for modeling $Q^*_D$ were proposed. The most recent
study reporting a method for the evaluation of $Q^*_D$ for Jupiter
Trojans was carried out by \cite{wong_differing_2014}. They simulated
the collisional evolution of the Jupiter Trojans by adjusting the
scaling law for $Q^*_D$ of main belt asteroids proposed by
\cite{durda_collisional_1998} to fit it to the low-diameter end of the
debiased size-frequency distribution of Trojans down to about 10
km. Because the composition of Trojans is believed to be different
from main belt asteroids and much smaller objects are studied in
this work, their approach is not applicable here. Instead, a scaling
law introduced by \cite{benz_catastrophic_1999}, which is valid down
to centimeter-sized objects, is used to determine $Q^*_D$:
\begin{equation}
 Q^*_D = Q_0\left({\frac{R_{tar}}{1\,\text{cm}}}\right)^a +
 B\rho\left({\frac{R_{tar}}{1\,\text{cm}}}\right)^b,
\label{equ:Q_D}
\end{equation}
where $R_{tar}$ and $\rho$ denote radius and density of the target
body, respectively. The advantage of this model is that it uses the
density as a parameter and that it is capable of describing $Q^*_D$
both in the strength regime (small objects) where fragmentation of the
target body is the dominant factor, and in the
gravity-dominated regime (large objects) where the target body is
fragmented and the fragments must also be dispersed in order to cause
a catastrophic disruption. Benz and Asphaug used a smoothed-particle
hydrodynamics code to simulate collisions of different materials and
impact velocities and provided $Q^*_D$ fits for each parameter
set. For icy material and an impact velocity of $v_{i}$ = 3 \uvel,
conditions closest to those in this work, the remaining parameters in
Equation \ref{equ:Q_D} are $Q_{0}$ = 1.6 J\,g\textsuperscript{-1}, $B$
= 1.2 $\times$ 10\textsuperscript{-7}
J\,cm\textsuperscript{3}\,g\textsuperscript{-2}, $a$ = -0.39, and $b$ =
1.26. To obtain a lower limit for the collisional lifetime, the
density of the weakest bodies (0.8 g/cm\textsuperscript{3}) was used
to evaluate $Q^*_D$. Knowing $Q^*_D$, the radius for an impactor
$R_{dis}$ capable of disrupting a target body of a certain radius
$R_{tar}$ can be calculated as in \cite{bottkejr_linking_2005},
\begin{equation}
  R_{dis} = \left({\frac{2Q^*_D}{v_i^2}}\right)^{1/3}R_{tar}.
\end{equation}
Finally, given the size distribution of the collisionally interacting
bodies, the collisional lifetime ($\tau_{dis}$) of an object with
radius $R_{tar}$ is \citep{farinella_meteorite_1998}
\begin{equation}
  \tau_{dis} = {\frac{1}{P_iR_{tar}^2N(>R_{dis})}},
\end{equation}
where $N(>R_{dis})$ is the number of objects with radii $R>R_{dis}$.
P\textsubscript{i} was set to the highest value reported, which is
7.79 10\textsuperscript{-18}
yr\textsuperscript{-1}km\textsuperscript{-2} for L\textsubscript{4} versus
L\textsubscript{4} Trojans \citep{davis_collisional_2002}. Other
populations such as short-period comets or Hildas also collisionally
interact with Jupiter Trojans \citep{delloro_updated_2001}. However,
the intrinsic impact probability between these populations and the
Trojans is orders of magnitude lower than the Trojan's mutual impact
probability, and the total number of objects of the collisionally
interacting populations is much smaller than the number of Trojans. Their influence on the collisional lifetime of Trojans is therefore negligible
for this experiment.

We know only very little about the size
distribution of $R<1$ km Jupiter Trojans, which we need to
calculate $N(\geq R_{dis})$. By observing and debiasing
L\textsubscript{4} Trojans using the 8.2 m Subaru Telescope at the Mauna Kea Observatory on Hawaii, \cite{yoshida_small_2017} found the size distribution
of objects between absolute magnitude $12<H<17.5$ (roughly corresponding to $1<R<10$ km) to follow a broken power
law. According to their work, the cumulative size distribution of
L\textsubscript{4} Trojans is
\begin{equation}
  \sum(H)=\begin{cases}
    10^{\alpha_1(H-H_0)}, & \text{for $H<H_b$}\\
    10^{\alpha_2H+(\alpha_1-\alpha_2)H_b-\alpha_1H_0}, & \text{for $H \geq H_b$},
  \end{cases}
\end{equation}
with the break magnitude $H_b = 13.56$ mag, corresponding to an approximate radius of 5 km, and power-law slopes $\alpha_1$ and $\alpha_2$ of 0.50 and 0.37 mag\textsuperscript{-1}, respectively.

Adding all this together, an estimation of the collisional lifetime of
Jupiter Trojans can be made. For the size ranges considered in this
experiment, the collisional lifetimes are given in Table
\ref{tab:tau_trojans}. Figure \ref{fig:trojan_size_tau} shows the
cumulative size distribution we used to estimate the number of targets
and impactors as well as the resulting estimate for the collisional
lifetime of L\textsubscript{4} Trojans. Although this estimation is
calculated based on the size distribution of L\textsubscript{4}
Trojans, it represents a lower limit for the collisional lifetime of
Jupiter Trojans in general because the L\textsubscript{5} Trojans are
expected to be fewer in numbers and thus should have even longer
collisional lifetimes than L\textsubscript{4} Trojans. The integration interval for investigating the influence of
the Yarkovsky effect was therefore set to 1 Gyr.

\begin{table}
\centering
\caption{Radii of target and projectile capable of causing a
  catastrophic disruption of the target body as well as the number
  projectiles in the L\textsubscript{4} cloud and the resulting
  collisional lifetime for the target body.}
\begin{tabular}{cccc}
\hline $R_{tar}$ [m] & $R_{dis}$\textsubscript{L\textsubscript{4}} [m]
& $N(R_{dis})$\textsubscript{L\textsubscript{4}} &
$\tau$L\textsubscript{L\textsubscript{4}} [Gyr]\\
\hline 10 & 0.20 & 2.55 $\times$ 10\textsuperscript{11} & 5.04\\
100 & 1.94 & 3.80 $\times$ 10\textsuperscript{9} & 3.37\\
1000 & 35.2 & 1.79 $\times$ 10\textsuperscript{7} & 7.17\\
10000 & 751 & 6.22 $\times$ 10\textsuperscript{4} & 20.1\\ \hline
\end{tabular}
\label{tab:tau_trojans}
\end{table}

We note that the collisional lifetime of small Trojans
calculated by \cite{de_elia_collisional_2007} is significantly shorter
than the estimate presented here. The authors considered
catastrophic disruptions as collisions where the objects are shattered
but not necessarily dispersed. Furthermore, they used a primordial
population of Trojans that was eight times more massive than the debiased
population found by \cite{jewitt_population_2000}. The focus in this
work is on the dynamical evolution of the already evolved population
rather than the collisional evolution of the primordial population. 
Even for a population containing ten times more
objects, the collisional lifetime for $R=100$ m objects would still be
in the order of several 100 Myr.

\subsection{Estimation of the libration amplitude}

Another quantity needed to interpret the results of this work is the
libration amplitude because it is an important indicator for the
long-term orbital stability of Trojans. The libration amplitudes of
Jupiter Trojans typically lies in the range between a few degrees to
about 35\degree. A convenient way to determine the libration amplitude
is to apply a frequency map analysis on the output of the numerical
integration in order to identify periodic variations in the orbital
elements and the resonant argument \citep{marzari_stability_2003}. To
estimate the libration amplitudes for the population of clones used in
our study, a simpler approach was employed. The libration amplitudes
were calculated from a short (200 kyr) integration. In order to remove
the short-period variations from the orbital elements, which would
modulate the libration, a digital low-pass filter was used. The
parameters were chosen such that periods shorter than the libration
period lie in the stop band, while the libration period (about 145
years or longer) lies in the pass band and is very little affected by
the filter. Using the filtered elements, we computed the amplitude of the
oscillation in semimajor axis ($d$) as the mean of the
maximum excursion in semimajor axis ($a - a_J$) in 2 kyr intervals
over the 200 kyr of integration,
\begin{equation}
a - a_J = d \sin\theta + \mathcal{O}({d}^2)
.\end{equation}
For our estimation, the higher order terms described by
$\mathcal{O}({d}^2$ are omitted.  Finally, the oscillation amplitude
in semimajor axis was translated into the libration amplitude ($D$) in
mean longitude using an approximation by \cite{erdi_long_1988},
  \begin{equation}
d = \sqrt{3 \mu} \, a_j D \approx 0.2783 D,
\label{equ:erdi}
\end{equation}
where $\mu$ is the ratio of the mass of Jupiter to the total mass of the
system.

In order to test this procedure, the libration amplitudes of the
numbered and multi-opposition Trojans were calculated and compared
with those available on
AstDyS\footnote{\href{https://newton.spacedys.com/~astdys2/propsynth/tro.syn}{https://newton.spacedys.com/~astdys2/propsynth/tro.syn}}. The
estimated amplitudes agree well. The average error is
0.61$\pm$0.01\degree, and only a few objects with very large libration
amplitudes or very long libration periods show larger errors.

\subsection{cuSwift}

The numerical experiments described in this paper place high demands
on computing power. In total, 131072 particles were integrated for 1
Gyr. It was decided to use the WHM integrator
\citep{wisdom_symplectic_1991} with an extension for the Yarkovsky
effect introduced by \cite{broz_yarkovsky_2006}. With the traditional
implementation, however, the calculations would either require
expensive supercomputer time, or the problem would need to be split
and distributed over an impractical number of workstation computers to
achieve results in acceptable computation time. In order to circumvent
these obstacles, the original integration methods were redesigned in
order to exploit the enormous computational power of multicore
processors and modern graphics processing units (GPUs). The newly
implemented integration methods are written in C/CUDA and follow the
implementations in
SWIFT\footnote{\href{https://www.boulder.swri.edu/~hal/swift.html}{https://www.boulder.swri.edu/\textasciitilde
    hal/swift.html}} and
Swifter\footnote{\href{http://www.boulder.swri.edu/swifter/}{http://www.boulder.swri.edu/swifter}}. To
ensure correctness and consistency, extensive tests were carried
out and test results were carefully compared with the original methods
\citep{hellmich_gpu_2017}. Our integration package, which we call
cuSwift, can produce output in the same format as Swifter. Simulations
with Swifter and the CPU version of cuSwift with the same initial
conditions produce exactly the same results. Because of slight differences
in the implementation of some mathematical operations on GPUs, results
obtained with the GPU version of cuSwift may differ from those
obtained with SWIFT/Swifter. However, the tests revealed that these
differences do not influence the overall accuracy. Some
mathematical expressions evaluated on the GPU produce even more
accurate values than on the CPU.

On a 2014 workstation PC, equipped with a 3.40 GHz Intel i7 Processor and two
Nvidia GeForce GTX Titan Black GPUs, the WHM implementation in cuSwift
outperforms the original implementation by a factor of about
100. Using that workstation, the main integrations of 98304 orbits
with and 32768 orbits without inclusion of the Yarkovsky effect over 1
Gyr took about 65 days in total.


\section{Results and discussion}

\begin{figure}
\centering
\includegraphics[width=\hsize]{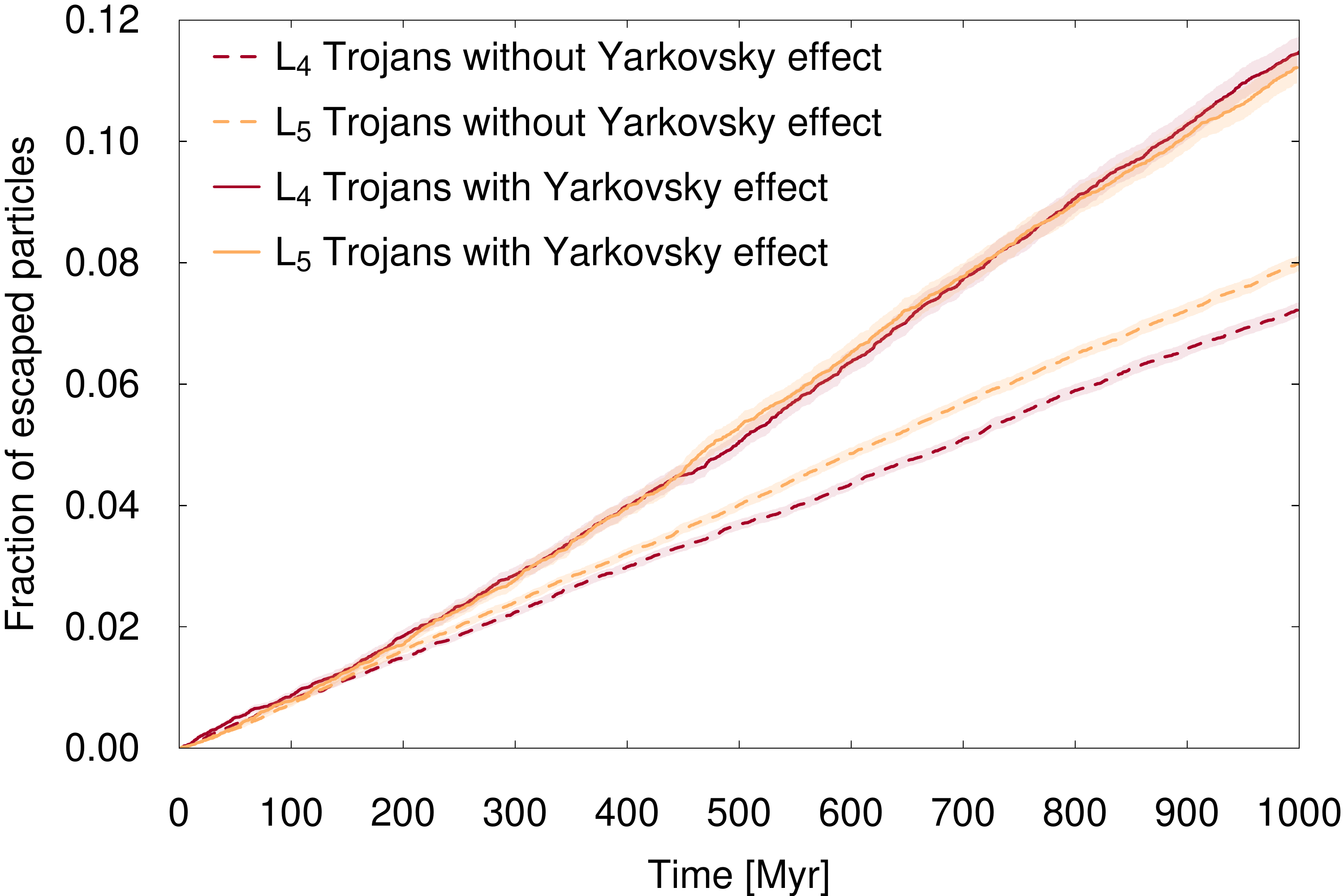}
\caption{Fraction of escaped particles with and without
  the Yarkovsky effect over 1 Gyr.}
\label{fig:escapes_yarko_v10}
\end{figure}

We ran a simulation including all planets from Earth to Neptune, with
the population of Trojans as described in Section \ref{sec:a_synth}
considered as massless particles. During the simulation we flagged all
particles when their heliocentric distance became smaller than 3
au, exceeded 8 au, or when the particles approached Jupiter closer than
twice the planet's Hill radius. Such particles were excluded from
further processing and considered escaped from the Trojan
region. Figure \ref{fig:escapes_yarko_v10} shows the fraction of
particles escaping from the Trojan clouds over 1 Gyr. Each line for
the non-Yarkovsky run corresponds to 49152 initial particles, while
each line for the Yarkovsky run represents a group of 16384 initial
particles with radii from 10 m to 10 km and with physical properties
as described in Table \ref{tab:props_trojans}. Transparent areas
represent the 1$\sigma$ confidence region, assuming the number of
ejected particles follows a Poisson distribution.

The Yarkovsky effect on our synthetic Trojan population
is striking. After 1 Gyr, significantly more particles were ejected
from the Trojan region than without the Yarkovsky effect.

\subsection{Influence of the physical properties}

\begin{figure*}
\centering
\includegraphics[width=1.0\textwidth]{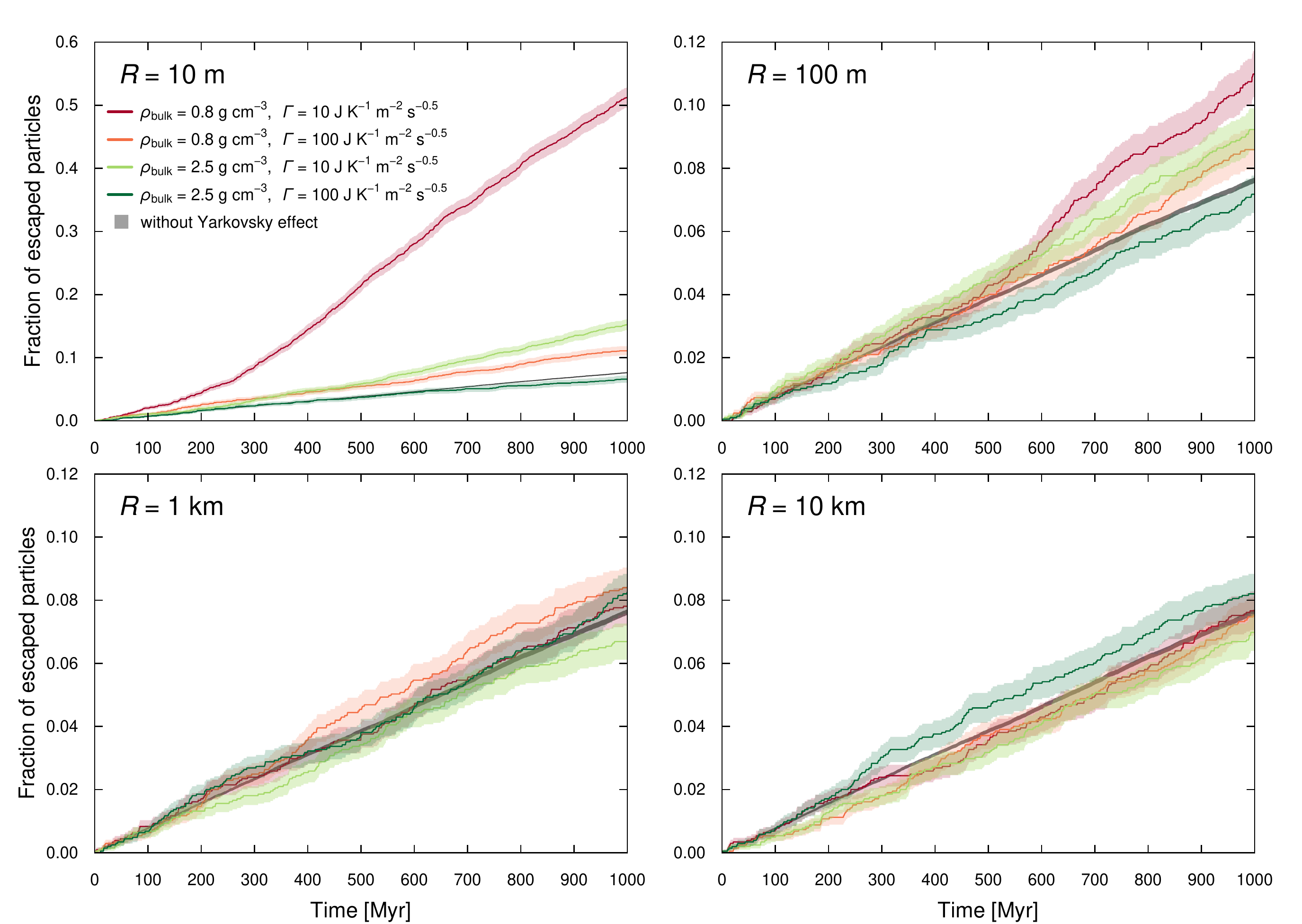}
\caption{Fraction of escaped particles for the different sizes and
  physical properties of the particles.}
\label{fig:discarded_categories}
\end{figure*}

In order to explore the influence of the physical properties
of the objects on the magnitude of the Yarkovsky-induced change in fraction of
ejected particles, we separately monitored the escaping fraction for
each category and size. Figure
\ref{fig:discarded_categories} shows that smaller objects are more likely to be
ejected from the Trojan region. Here, each line corresponds to a group
of 2048 initial particles, and the colored transparent areas in the
plots represent the 1$\sigma$ confidence region for each category and
size. The gray area in each plot is the confidence region for the
non-Yarkovsky run. Because the initial number of particles for each
category and size in the Yarkovsky run was 2048 and the number of
particles in the non-Yarkovsky run was 98304, the 1$\sigma$
confidence region for the non-Yarkovsky run is accordingly narrower
than the region corresponding to the subpopulations of the Yarkovsky
run. While for objects of $R=1$ km and $R=10$ km no significant
difference in the escaping fractions due to the Yarkovsky effect was
detected, for objects with radii of 10 and 100 m, low thermal inertia, and low
density, more than 7 and 1.5 times as many particles were ejected
over the integration interval of 1 Gyr of the
Yarkovsky run than during the non-Yarkovsky run, respectively. Moreover, the fraction of
escaped particles with low thermal inertia and high density and
the fraction for the objects with high thermal inertia and low density are clearly above
the confidence region of the non-Yarkovsky run. The consequences of
this finding reach far. The efficient removal of small Trojans
through the Yarkovsky effect implies that below a radius of 100 m these
objects are increasingly depleted, and might even be entirely removed
at very small sizes. As a consequence, we expect that the
size-frequency distribution of Trojans shows a shallower slope at
smaller radii, with a turning point between $R=100$ m and $R=1$ km,
just below the size regime observed by
\nopagebreak\cite{yoshida_small_2017}. The shape of this
size-frequency distribution is expected to be imprinted on the cratering
record of Trojans.

An interesting feature for the particles with $R=10$ m, low density, and low thermal
inertia is that the rate of ejected particles per unit time
increases after about the first 250 Myr. If the initial orbital
distribution were in a steady state, as can be assumed for an
evolved population like the Jupiter Trojans, we would expect the
escape rates to be constant over time for each size and category of
physical properties. A possible explanation for the rate change is
that the initial orbital distribution of the test particles for this
experiment reflects the distribution of the currently known Jupiter
Trojans. This distribution is based on objects much larger than 1 km
in radius, which are less strongly influenced by the Yarkovsky effect. Most of
them have very long dynamical lifetimes (fewer than 5\% of all
particles were ejected during the non-Yarkovsky run), that is, they are
on very stable orbits. The reason for the suddenly increasing escape
rate would be that it takes some time until the Yarkovsky force moves
the much smaller particles we considered in this experiment from their
stable initial orbits to less stable regions from which they are
eventually ejected. For small particles this is expected to occur faster than
for larger particles because they are more strongly affected by the
Yarkovsky force. The same effect becomes visible for the $R=100$ m
objects. Except for objects with high thermal inertia and high density,
the escape rate increases after about 500 Myr of integration. The
change in slope implies that the initial conditions for this
experiment may not represent the steady-state orbital distribution for
Trojans smaller than 1 km in radius and suggests that the actual
number of escaped particles could be higher. It is understood that the
details of the change in slope will depend on the actual distribution
of the physical and rotational properties of the particles, especially
obliquity. However, we expect the general trend to remain.

\subsection{\texorpdfstring{Ratio between escaped L\textsubscript{4} and L\textsubscript{5} Trojans}{...}}

The asymmetry between the two Trojan clouds is observed on the known
population, which consists of objects with radii of tens to some hundred
kilometers. Because these bodies are hardly influenced by the Yarkovsky
effect, the latter cannot explain the asymmetry at large
sizes. Furthermore, \cite{hou_dynamics_2016} demonstrated that in the case
of a restricted three-body problem containing the Sun, Jupiter, and a
Trojan asteroid, the influence of the Yarkovsky force is slightly
different for the two Lagrangian regions, suggesting that their
long-term orbital stability might also differ. It is
therefore meaningful to examine whether the effect may influence the ratio of
escaping particles between L\textsubscript{5} and L\textsubscript{4}
for small objects in a more realistic scenario where perturbations
from the other planets are considered as well. In Figure
\ref{fig:escapes_yarko_v10} the escaping fractions are separately
monitored for the two clouds. For the non-Yarkovsky run, the ratio
between escaped L\textsubscript{5} and L\textsubscript{4} particles
after 1 Gyr was 1.106$\pm$0.026, which shows that L\textsubscript{5}
objects are indeed depleted at a faster pace than L\textsubscript{4}
Trojans, although not fast enough to explain the currently observed
asymmetry. This value, which appears to be roughly constant as the
system evolves, agrees well with the ratio of 1.11$\pm$0.08 that was
computed using the escape rates reported by
\cite{di_sisto_dynamical_2019}, who integrated the numbered Jupiter
Trojans for 4.5 Gyr without the Yarkovsky effect.

\begin{figure*}
\centering
\includegraphics[width=1.0\textwidth]{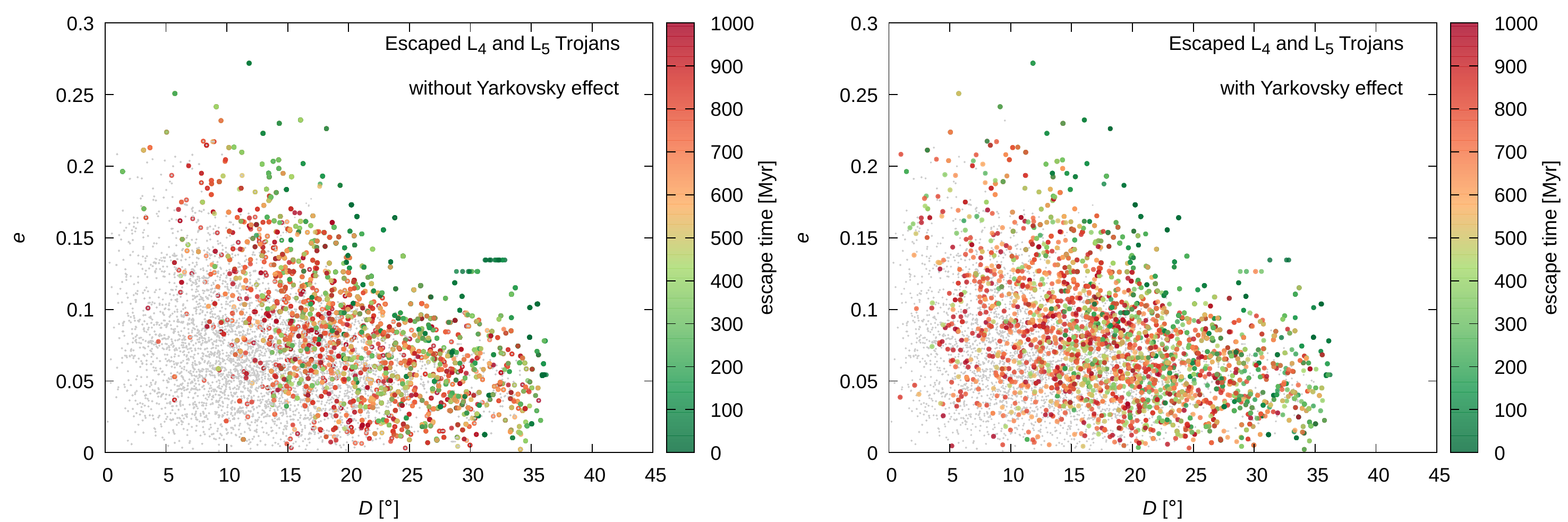}
\caption{Initial amplitude and escape time for the escaped particles
  during the non-Yarkovsky run (left panel) and the Yarkovsky run
  (right panel). The small gray dots represent the survivors.}
\label{fig:discarded_libamps}
\end{figure*}

When the Yarkovsky effect is activated, the overall fraction of the
escaping particles from L\textsubscript{4} and L\textsubscript{5} is
virtually identical. Figure \ref{fig:escapes_yarko_v10} shows no trend
of more particles escaping from either of the two clouds. Grading the
particles according to radius suggests that the ratio of escaped
particles between L\textsubscript{5} and L\textsubscript{4} decreases
with size, although the uncertainty is too large to allow a significant
conclusion. In order to verify whether the Yarkovsky effect indeed
causes the L\textsubscript{5}/L\textsubscript{4} escape ratio to
decrease with size, an additional integration incorporating 32768 R =
10 m particles with low density and low thermal inertia (category 1 in
Table \ref{tab:props_trojans}) over 1 Gyr was performed. For these
bodies, the influence of the Yarkovsky force is greatest. The
L\textsubscript{5}/L\textsubscript{4} escape ratio for this run is
0.985$\pm$0.016, meaning that the L\textsubscript{5} and
L\textsubscript{4} escape rates are indistinguishable with our sample
size.  The fact that the L\textsubscript{5}/L\textsubscript{4} escape
ratio becomes unity for the Yarkovsky run including only bodies with $R$ = 10 m,
low density, and low thermal inertia supports the notion that
the Yarkovsky effect causes the orbital distribution in the two clouds
to become more homogeneous.

\subsection{Libration amplitude}

\begin{figure*}
\centering
\includegraphics[width=1.0\textwidth]{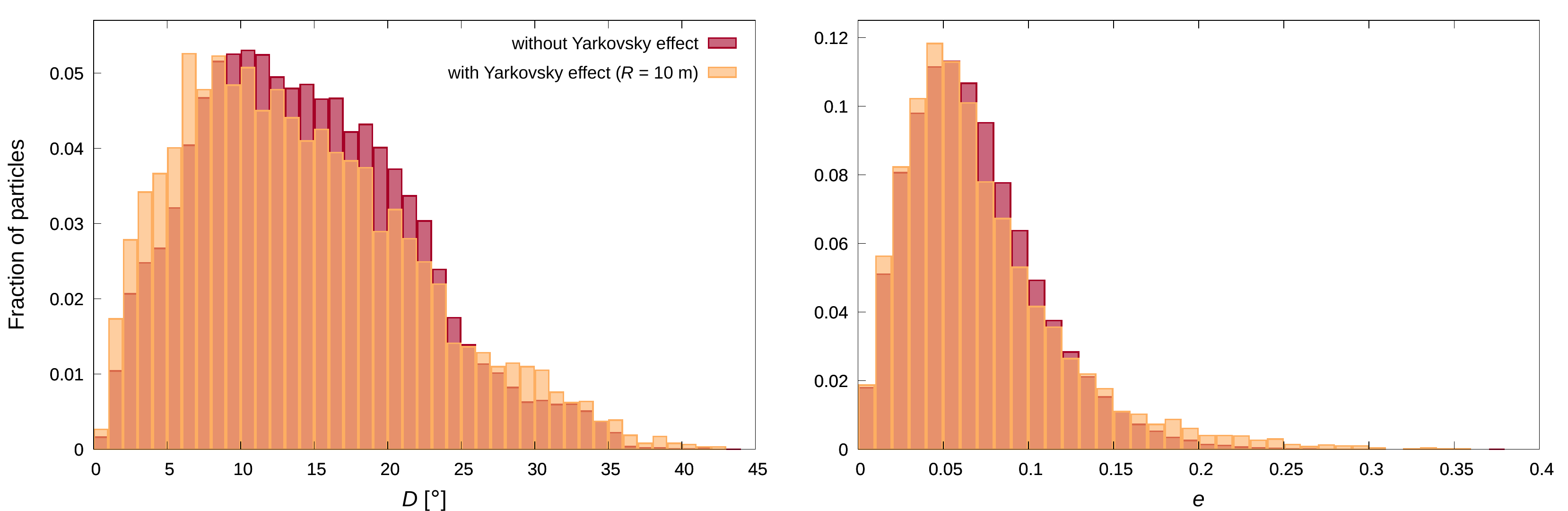}
\caption{Comparison of the histograms of libration amplitudes (left)
  and eccentricities (right) of particles that remained in the Trojan
  region over the complete integration time span in the Yarkovsky and
  non-Yarkovsky run after 1 Gyr (10 m radius objects only).}
\label{fig:libamp_ecc_hist_S1}
\end{figure*}

To understand better how the Yarkovsky force changes the
long-term orbital evolution of the Trojans, the initial libration amplitudes and
eccentricities of the particles ejected during both runs were
compared. In general, orbits with a low libration amplitude,
eccentricity, and inclination tend to be more stable. Figure \ref{fig:discarded_libamps} shows that objects with high eccentricity
or libration amplitude were ejected early during the simulation, while
those with lower eccentricity or libration amplitude remained for a
longer time span. The figure also shows that more particles were
ejected that were initially located in the low libration amplitude
regime during the Yarkovsky run as during the non-Yarkovsky run. The
average initial libration amplitude for the particles ejected during
the non-Yarkovsky run was 21.75$\pm$0.12\degree\ for
L\textsubscript{4} and 20.85$\pm$0.10\degree\ for
L\textsubscript{5} Trojans. For the Yarkovsky run, these values
slightly decreased to 20.19$\pm$0.17\degree\ and
19.51$\pm$0.17\degree\ for L\textsubscript{4} and L\textsubscript{5}
Trojans, respectively. Although the difference is not very large, it
is significant, and it indicates that the Yarkovsky effect causes objects
that initially were on stable orbits to leave the Trojan region. This
trend should also be reflected in the distribution of the libration
amplitudes and eccentricities of the survivors at the end of the
integration. If the initial population is not in steady state, we
would expect more objects with higher amplitudes and eccentricities
after 1 Gyr integration including the Yarkovsky effect than there are
in the non-Yarkovsky run. Thus, the libration amplitudes of the
surviving particles at the end of each simulation were computed and
compared. Figure \ref{fig:libamp_ecc_hist_S1} shows histograms of
libration amplitude and eccentricity at the end of the
integration. Because the Yarkovsky force has the strongest effect on objects with
10 m radius, only these objects are plotted as histograms for
the Yarkovsky run. There clearly is an abundance of large libration
amplitude and high-eccentricity orbits in the Yarkovsky run. The
differences decrease with increasing size. While for $R=10$ m to $R=1$
km objects, a two-dimensional Kolmogorov-Smirnoff non-parametric test
rejected the null-hypothesis (that $D$ and $e$ after the Yarkovsky run
and the non-Yarkovsky run follow the same distribution) with a
significance level lower than 0.01, the test was unable to distinguish
between distributions for 10 km objects.

The Yarkovsky effect did not increase the libration amplitudes and
eccentricities for all objects: Figure \ref{fig:libamp_ecc_hist_S1}
clearly shows that there is also an abundance of low amplitude and low-eccentricity objects after 1 Gyr of integration, which indicates that
the Yarkovsky effect tends to increase the dynamical lifetime for some
of the small objects. The same trend was also present when we considered either of the Trojan clouds separately, suggesting that both clouds are
affected in the same way. The drift in semimajor axis caused by the
diurnal component of the Yarkovsky force depends on the sense of the
object's rotation. In general, the Yarkovsky effect leads to an
increase in the semimajor axis of prograde rotators and
decreases the semimajor axis of retrograde rotators.

Trojans are in a 1:1 mean-motion resonance with Jupiter, and their
semimajor axes are coupled with that of the gas giant. For these
objects, the Yarkovsky effect causes a secular change in the
oscillation amplitude of the semimajor axis. The libration cycle of
Trojans typically lasts about 150 yr. An object spends about half of
that time on an orbit with a semimajor axis smaller than that of
Jupiter and the other half on an orbit with a semimajor axis larger
than that of the gas giant. Assuming prograde rotation, a Trojan would
be pushed toward the libration center during the first half of the
libration cycle and dragged away from it during the second
half. Because the magnitude of the Yarkovsky force changes
proportionally to $1/r$, where $r$ is the distance to the Sun, the
force pushing the object toward the libration center is slightly
stronger than that dragging it away from it, which results in a
secular decrease of the oscillation amplitude. Retrograde rotation
would result in the opposite effect. This mechanism was also shown by
\cite{wang_dynamical_2017} in an idealized experiment employing
the circular restricted three-body and a simplified model for the
Yarkovsky effect. They found that under the influence of the Yarkovsky
force, prograde rotators experience an exponential decrease of the
oscillation amplitude and that the amplitude is exponentially increased
for retrograde rotators. The eccentricity changes in the opposite
direction (decreasing for retrograde rotators and increasing for
prograde rotators), but the inclination remains unaffected. They
further showed that when the gravitational influence of
Saturn is included, the oscillation in semimajor axis for prograde
rotators at some point also increases because the eccentricity becomes so high that the
influence of Saturn dominates the orbital evolution.

\begin{figure*}
\centering
\includegraphics[width=1.0\textwidth]{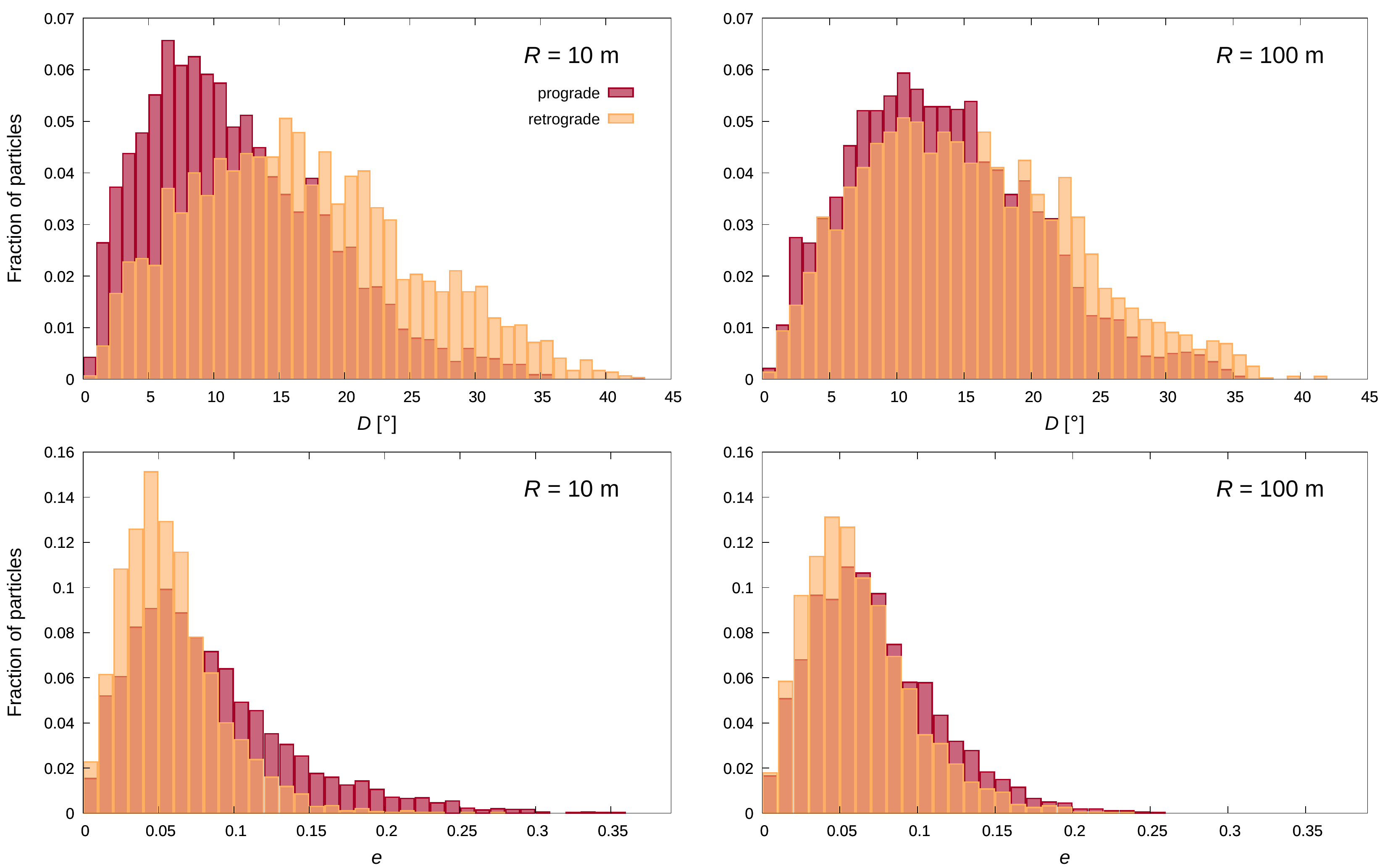}
\caption{Distribution of libration amplitudes and eccentricities of objectse with 10
  and 100 m radii that survive 1 Gyr of integration under the
  influence of the Yarkovsky force.}
\label{fig:libamp_ecc_hist_obl_S1}
\end{figure*}

In order to verify if indeed a prograde rotation causes the observed
abundance of low-amplitude objects in our experiment, libration
amplitude and eccentricity were analyzed separately for prograde and
retrograde rotators. Figure \ref{fig:libamp_ecc_hist_obl_S1} shows
histograms of libration amplitude and eccentricity for prograde versus
retrograde rotators after 1 Gyr of integration for objects of $R=10$ m
and $R=100$ m. While there are more prograde rotators in the low
libration amplitude regime (implying longer dynamical lifetime), there
are also more prograde rotators in the high-eccentricity regime
(implying shorter dynamical lifetime). The average values are listed
in Table \ref{tab:avg_D_e_rot}.

\begin{table}
\centering
\caption{Average values for libration amplitudes after 1 Gyr integration considering the Yarkovsky effect.}
\begin{tabular}{cccc}
\hline
$R$ & & $\bar{D}$ [\degree] & $\bar{e}$ \\
\hline
\multirow{ 2}{*}{10 m} & prograde & 11.84 $\pm$ 0.12 & 0.083 $\pm$ 0.001 \\
& retrograde & 16.75 $\pm$ 0.15 & 0.058 $\pm$ 0.001 \\
\hline
\multirow{ 2}{*}{100 m} & prograde & 13.56 $\pm$ 0.11 & 0.072 $\pm$ 0.001 \\
& retrograde & 15.43 $\pm$ 0.13 & 0.063 $\pm$ 0.001 \\
\hline
\multirow{ 2}{*}{1 km} & prograde & 14.12 $\pm$ 0.12 & 0.068 $\pm$ 0.001 \\
& retrograde & 14.44 $\pm$ 0.12 & 0.066 $\pm$ 0.001\\
\hline
\multirow{ 2}{*}{10km} & prograde & 14.62 $\pm$ 0.12 & 0.067 $\pm$ 0.001 \\
& retrograde & 14.44 $\pm$ 0.12 & 0.066 $\pm$ 0.001 \\
\hline
\end{tabular}
\label{tab:avg_D_e_rot}
\end{table}

We again applied the two-dimensional Kolmogorov-Smirnoff test. This showed
that libration amplitudes and eccentricities of prograde and
retrograde rotator objects with $R=10$ m and $R=100$ m do not follow
the same distribution with a significance level lower than
0.001. For larger objects, the test was unable to distinguish between the
two distributions. The fraction of escaping
particles also shows that the destabilizing influence of the
Yarkovsky effect is greater for retrograde than for prograde
rotators. In total, there were 8272 prograde and 8112 retrograde
objects with $R=10$ m and $R=100$ m of which after 1 Gyr, 929
(11.2\%) and 1533 (18.9\%) were ejected, respectively. The fact that
at the end of the simulation we found more prograde rotators with both
small libration amplitude and high eccentricity can be explained as
follows: during the phase when the eccentricity is still low, the
libration amplitude decreases, but when the eccentricity becomes so high
that the gravitational influence of Saturn dominates, the libration
amplitude increases rather fast, which eventually causes the particles to be ejected from the Trojan region.

The finding that the Yarkovsky effect causes small retrograde rotators
to be ejected faster than prograde rotators is expected to be reflected in the
obliquity distribution of the small Jupiter Trojans, which currently
is unknown, however. The influence of the Yarkovsky force
on the obliquity distribution of Jupiter Trojans might therefore be confirmed by
observations in the future.

We found no impact of the Yarkovsky effect on the inclination. The distributions of the inclination resulting from the two simulations for each cloud and
size do not differ significantly.

\subsection{Caveats}

In order to assess the significance of
the results of this work, we  list three main caveats. First of all, the physical and thermal properties of the
test particles were set to values obtained for rather large objects
and might be different for smaller bodies considered in this
experiment. The parameters for the scaling law by
\cite{benz_catastrophic_1999} we used to determine $Q^*_D$ are valid for
icy bodies and an impact velocity of 3 \uvel. However, there is no
evidence so far that Trojans are indeed icy bodies, and the impact
velocity lies between 4.5 and 5 \uvel
\citep{davis_collisional_2002}. Second, the particles were assumed
to be perfect spheres, as for most studies modeling the dynamical
evolution of a huge number of bodies under the Yarkovsky effect. This
simplification might have an influence on the magnitude of the
Yarkovsky effect, especially for small bodies, which can have very
irregular shapes. Moreover, the YORP effect, which changes the rotational
properties, was not considered in this study. Over long timescales, the YORP effect tends to drive the obliquity toward
0\degree\ and 180\degree\ for prograde and retrograde rotators,
respectively \citep{vokrouhlicky_vector_2003}, resulting in a bimodal
rather than uniform distribution of spin vector
orientations. Obliquities close to 0 or 180\degree\ cause the
diurnal component of the Yarkovsky effect to become stronger, and
because for Trojans this component is dominant, a bimodal distribution
of spin vectors would lead to a stronger Yarkovsky
effect. Last but not least, the smaller the objects, the greater their number, and consequently, the higher the frequency of their collisions that, when not disruptive, change the spin-axis
orientation and spin rate of the object. This in turn results in a random Yarkovsky-drift
direction over very long timescales.

\section{Conclusions}

By simulating the orbital evolution of a synthetic population of
Trojan asteroids under the influence of the Yarkovsky effect over 1
Gyr, we obtained the following results: 1) We proved that small
objects up to the size of 1 km in radius can be affected by the
Yarkovsky force, depending on their thermo-physical
properties. Compared with a purely gravitational orbital evolution,
the Yarkovsky effect leads to a significant shortening of the
residence time. 2) Owing to the Yarkovsky effect, objects with $R \leq
1$ km are expected to exhibit a distinct orbital distribution from the
currently observable population of larger objects. In particular,
purely gravitational models predict that more particles are ejected from
L\textsubscript{5} than from L\textsubscript{4}. When the
Yarkovsky force is considered, the escape rates from both clouds become equal,
suggesting a more homogeneous orbital distribution in both clouds for
small objects. 3) The experiment thereby confirmed that the difference
in the number of Trojans that escape from L\textsubscript{4} and
L\textsubscript{5} cannot be explained by the Yarkovsky effect and
must be caused by the different preexisting orbital distributions of
each group.  4) The Yarkovsky effect causes the residence time in
Trojan orbits to decrease with decreasing object radius. Therefore
fewer Trojans with $R \leq 1$ km are expected to exist than is predicted by
purely dynamical and collisional models.  5) The residence time of small
objects in the Trojan region depends on their sense of rotation. For
retrograde rotators, the Yarkovsky effect shortens the residence time
in the Trojan clouds because the libration amplitude increases.
For prograde rotators, the permanence time is also reduced, but to
a lesser extent, as a result of the increase in eccentricity.

Some of our findings implicitly assume that small Trojans have the
same size-frequency distribution slope as larger bodies. To which
extent this assumption is justified is currently unknown. The recently
selected Lucy space mission \nopagebreak{\citep{levison_lucy_2017}}
may provide valuable insights into this. During its 12-year mission,
the spacecraft will fly by five Trojans and deliver high-resolution
images of their surfaces. From the cratering records on these bodies
it will be possible to measure the size-frequency distribution of the
impacting bodies down to a size of~5 m, which may reveal the
signature of the Yarkovsky effect on Trojans that we found here.

\begin{acknowledgements}
The authors are grateful to Tilman Spohn for the continuous support and to Jürgen Oberst and Enrico Mai for the useful discussions. The authors are indebted to David Nesvorny and Tim Holt for their thorough review, which contributed to improving this paper.
\end{acknowledgements}

\bibliographystyle{aa} 
\bibliography{trojans} 

\end{document}